# Laws of volume elasticity of the physical processes and the parameter effect.

**P.I. Polyakov**

Mining Processes Physics Institute of NAS of Ukraine, 72 R.Luxemburg Str., Donetsk 83114, Ukraine

E-mail: poljakov@mail.fti.ac.donetsk.ua

**Abstract.** We consider the mechanism of elastic strains and stresses as the main controlling factor of structure change under the influence of temperature, magnetic field, hydrostatic pressure. We should take into account that the energy of elastic deformation is commensurate to the energy of electric interactions and that is much higher than the rest of the bonds of lower energy value. Besides, the energy elastic stresses are of long range, so it forms the linearity in magnetization and bulk change. These regularities requires a fundamental understanding of the laws of interaction with respect to accepted interpretation of quantum mechanical forces of short range that are attributes of magnetism formation. Due to the high sensitivity of electronic and resonance properties with respect to small changes of the structure, we were able to define the direct relation between elastic stresses and field-frequency dependencies, as well as to analyze the evolution of the dynamics of phase transitions and phase states. A cycle of studies of the influence of hydrostatic pressure on the resonance properties are presented also. The analysis of the effect of magnetic, magneto-elastic and elastic energy allowed us to define the combinations of magneto-elastic interactions.

The role of elastic stresses in the linear changes of the magnetostriction, magnetization, magnetoelasticity of single-crystal magnetic semiconductors is described in details.

## 1. Introduction

Physical processes of magnet-containing structures are supplemented with established laws, critical states, properties, that gave rise to fundamental experimental studies. We suggest analysis and consideration of evolution of a solid through the processes of bulk elastic deforming stresses transformed by the impact of temperature (T), hydrostatic pressure (P) and magnetic field (H) that form structural phase transition determining the state of properties, phenomena and effects. Physical models an not be foreseen without detailed study of processes that form the structure and the dynamics of interaction of electron bonds. We believe that high temperature structure formation process should be taken as a base of with the succeeding heat sink to the extremal low-temperature states that are mostly studied. That is the area where the application of theoretical models is done without accounting for the processes of volume change and the causing role of elastic deforming stresses.

The evolution of the science requires the use of more sensitive methods of investigation that divided the processes of cognition into limited narrow areas where it was difficult to formulate some general laws. The analysis and generalization of a wide set of analogies and results of experimental studies in well-studied metals[1], semiconductors[2,3], dielectrics[4], superconductors[5] allowed to put a question about the commonality of the mechanisms of influence of thermodynamic parameters (T-H-P) on the structure and the properties[9].

**2. General conceptions**

While considering the properties of solids as a set of incoming chemical elements that form the lattice structure, dynamical lattice changes are revealed as the evolution under T-P-H influence at the formation of the structural phase transitions and phase properties. The conditions determining the locations of atoms at lattice sites are high-temperature physical and chemical processes of the structure formation. Numerous further studies of physical processes and models are connected with the methods of the realization and the search for the major driving force (energy) determining mechanisms of the changes, inner interactions, structure, phase transitions and properties where the processes of structure formation and elasticity are not often taken into account.

As shown by the specific dynamics of the changes in the course of the T-P-H effect, we can establish and separate the causing role of long-order anisotropic elastic deformation stresses that form structural phase transitions of displacement and rupture.

We consider the models of the nature of interactions binding atoms and ions at the sites of the lattice structure where the determining role is played by free and valency electrons. Those are peripheral electrons with the energy lower by two or more orders that bind a set of atoms and ions within a lattice in the course of structure formation. Being transformed by heat sink, the process thermal compression changes the ratio of all energetical states of electron interactions and forms both the conducting and magnetic properties.

The estimations allowed us to suppose that the energy of elastic stresses is more comparable with the energy of exchange interactions and is the major energy parameter determining the bonds and states of all interactions to be less important. As a result, the elastic deformation stresses influence the structure changes and determine the commonality of mechanisms of the effect of the parameters (T-P-H) on the evolution and formation of structural phase transitions, critical states and properties. That fact excludes the existence of an equilibrium state of the studied object. Thus, the influence of the energy of elastic stresses on the interaction energy modifies the energy of electron bond between atoms of the structure. We should keep in mind that the energy of dense bonds between the structure sites at definite conditions forms the magnetic properties, and the energy of free electrons is related with the state of conductivity.

We should take into account that the energy of elastic deforming stresses is seen in linear regularities of resistivity, magnetization and striction.

**3. Bulk elasticity effects in different types of magnetic and conducting media.**
   *Conventional metals.*

Such a position simplifies understanding of the main laws of interactions with respect to the analysis of the experimental results of concrete physical processes.

As an example, we should draw attention to the works of Kapitsa [1] dealing with the bulk elasticity and the conductivity in metals. As shown by analysis of about 36 metals, the characteristic linear changes of the conductivity are stable within a wide range of field H. The typical examples are presented on the dependence of specific resistance of copper (Fig.1) at fixed $T_1$=86 K (1), $T_2$=63 K (2), $T_3$=20 K (3) with H varied from 0 to 300 kOe. The structure of Cu type possesses the magnetostriction properties in the considered temperature range in magnetic field and reacts on the influence of T and H parameters by the deforming violation of lattices. In the first place, it results in linear change of conductivity (Fig.1).

*Magnet-containing semiconductors and dielectrics.*

Consecutive analysis of the effect of elastic deformation stresses on the formation of structural phase transition and magnetic properties of the mentioned magnet-containing media raised a question, how the properties of conductivity behave at any changes of structure influenced by elastic deformation stresses.

The demonstrative are results of [3, 9] considering mechanisms involved in formation of the dynamics of resistive properties in magnetic

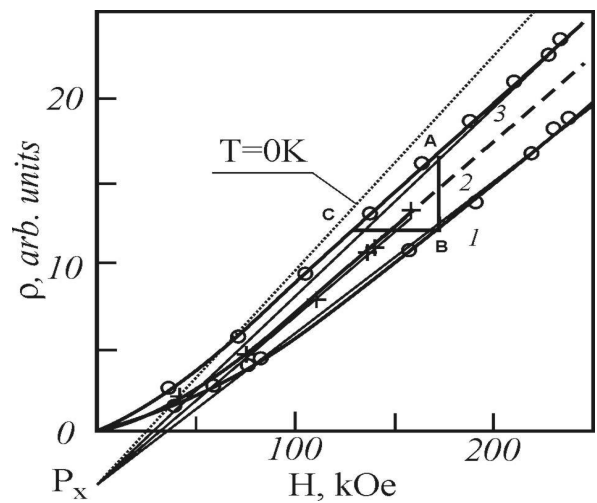

**Figure 1. The magnetoresistivity in copper at fixed temperatures: $T_1$=86 K, $T_2$=63 K, $T_3$=20 K.**

semiconductors. The authors studied the behavior of resistivity in a bulk ceramic sample $La_{0.56}Ca_{0.24}Mn_{1.2}O_3$ as a function of three thermodynamic parameters T-P-H (Fig.2). The originality of this paper is that three thermodynamical parameters T-P-H are studied simultaneously. The merit of such a method of the study of resistivity is the most apparent for the mechanisms of realization of physical processes. The changes taking place in non-magnetic chamber of high pressure at hydrostatic conditions are related to the process of structure transformations, and optimal ration of the volume of the sample and the liquid is 1:100 to within P~0.1%. The selected magnetic structure and its magnetic component are sensitive to the parameter influence with respect to the structural changes of interatomic bonds. Fig.6a presents the conductivity at the varied thermodynamic parameters. The decrease of the temperature (heat sink) results in the change of resistivity in semiconducting and metallic phase separated by the PT. The succeeding influence of elastic deformation in the form of P (pressure effect) and magnetic field H with accounting for the critical point $P_{x1}$ in the whole temperature range reduces the resistivity and changes the location of $T_{mc}(H)$ and $T_{mc}(P)$. Thus, magnetoelastic and baroelastic parameters influence interatomic bonds of the structure through elastic stresses that is revealed as critical phenomena in the form of resistivity jump

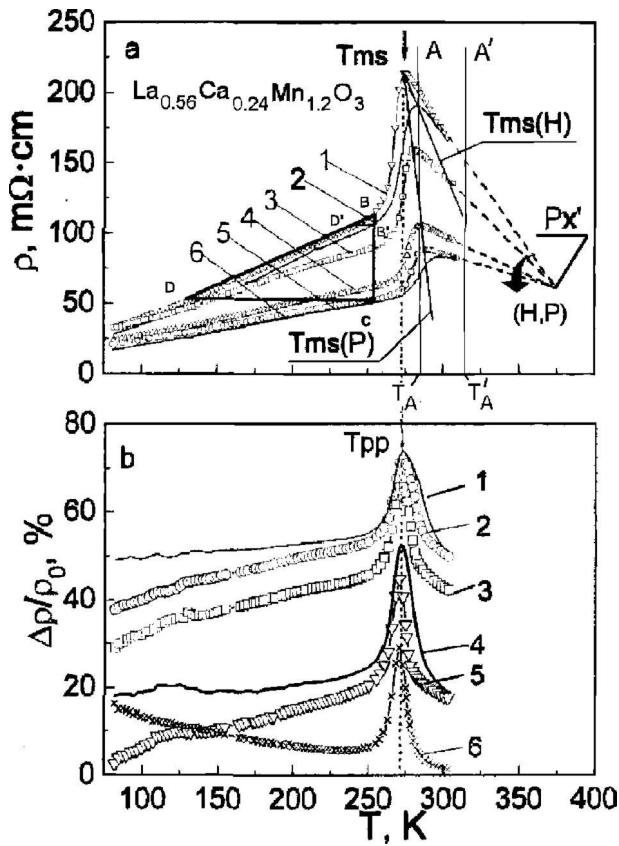

Figure 2. a) Temperature dependence of resistivity of $La_{0.56}Ca_{0.24}Mn_{1.2}O_3$ ceramic sample: 1 - P = 0 kbar; 2 - P = 0, H = 8 kOe; 3 - P = 6 kbar; 4 - P = 12 kbar; 5 - P = 18 kbar; 6 - P = 18 kbar, H = 8 kOe;

b) Temperature dependence of baroresistive, baromagnetoresisitive, magnetoresistive effects: 1 - H = 8 kOe, P = 18 kbar; 2 - P=0, H = 8 kOe; 3 - P = 12 kbar; 4 - P = 6 kbar; 5 - P = 6 kbar H = 8 kOe; 6 - H = 8 kOe.

The dynamics of their changes is affected by the PT and the differences of anisotropies of elasticity and magnetoelasticity. The character of the process depends on the elastic ductility and the level of the transformed stresses.

The established physical processes are repeated in other results of the investigations that is characteristics of the structure changes at the rate of the energies of elastic volume stresses. That is external impact through the elasticity that transforms the interaction of inner energies making changes of the structure and forming structural PT and phase states. In addition, we shall establish the role of cooling-heating effects of P and H. The ratio of the parameters T-P-H is 6.8K- 0.1GPa – 2.8kOe. These values for the given sample are obtained basing on the linear regularities of the experimental curves. Let us consider the effects using the triangle BCd and $AC_1d_1$ (Fig.2). We trace the position of B point at the fixed P that is 1.81 Pa. The resistivity is reduced analogously to D with the same properties that is similar to the reduction from C with the corresponding heat sink about 120 K. The inverse effect is the taking down of P to zero corresponds to the increase of resistivity at the rising temperature. All these changes result in structural transformations. The analogous result gives the influence of the magnetic field H~8 kOe with the shift of C point to B` similar to the temperature reduction of 20 K. The inverse effect with the absent magnetic field H=0 is followed by the increase of the resistivity analogously to the increased T.

Taking into account high sensitivity of electronic properties to insignificant structure changes, we confirm the supposition concerning adequate influence of thermodynamic parameters on the structural changes through the laws of volume stresses. Thus, it is very important to include into the model both structural changes and levels of elastic deformation stresses, i.e. the elastic energy transformed by the parameters in physical processes.

As formulated, there is an important regularity in such processes, i.e. conductivity redistribution because of the relaxation of inner stresses during volume reduction. As the thermo-elastic expansion is associated with considerable inner stresses, the influence of elastic and magnetoelastic compression induced by the back-pressure diminishes the inner stress. The process clarifies that any volume change due to thermo-elastic, baro-elastic, magneto-elastic compression relates directly to the inner-stress reduction in the system, while a volume decrease is directly related to the enhancement of the influence of energy of atom-electron coupling, by definition.

On the dependence (Fig. 2), the denoted important factor of a nearly double jump in the region of PT, where the phase states are separated, implies that changes have occurred already in the semiconducting phase. Such changes of properties are based on the presence of structural phase transition in a reversible physical process. In this case, the regularity is the equality of temperature maxima of thermobaroresistive, thermomagnetoresistive and thermobaromagnetoresistive effects TPP=Tmc that corresponds to the temperature of the structural phase transition. It can be consequently stated that the regularity TPP-const is the law of conformity conservation for elasticity and magnetoelasticity anisotropies.

*Magnetostrictionand magnetization analogies.*

Numerous studies of copper chloride dihydrate $CuCl_2 \cdot 2H_2O$ [4, 9] were taken as a basis for drawing analogies and correspondences in mechanisms and regularities of changes of structural phase transitions and properties of the mentioned multi-component magnet-containing systems. The interrelation of elastic mechanisms was established for the joint thermodynamic influence of P-T-H on multi-component magnet-containing structures.

Using the method of analogies and comparisons [3], we have brought to conformity the dependencies of magnetostriction in $LaMnO_3$ and magnetization in $CuCl_2 \cdot 2H_2O$. We have emphasized the causing role of thermoelasticity and magnetoelasticity in linear regularities of the properties (Figs. 3, 4a). We should note that the same features illustrate one more important regularity of identical effect of T and H on both magnetostriction dependencies (Fig. 3) before the phase transition in $LaMnO_3$ and magnetization (Fig. 4a) after the phase transition in $CuCl_2 \cdot 2H_2O$. We state that the separated areas have the same mechanisms of influence and belong to the conducting (metallic) phase of both the structures in the region where changes of the properties under T and H are antiphase. The next important result is established in magnetization behavior at the initial stage (Fig. 4a). We have singled out stable tendency of the change of properties with respect to the critical point $PP_x$ at similar T and H influence and the regularity of the motion of the critical line $T_p(H)$ of the temperature-field dependence was established that gave us an opportunity to demonstrate the dynamics of structural phase transition of the first type at T=0 K with the definition of the critical point $P(T_p=0$ K, $H_p=5.5$ kOe).

These results form new approaches to the analysis of phase diagrams [1, 3], low-frequency properties both at 0 K and at selected temperatures in $CuCl_2 \cdot 2H_2O$. New methods of analysis using approximations of the corresponding curves based on the elastic linearity don't contradict to the logic of the studied reversible processes and allow to define the position of a structural phase transition of the second type at 0K both in $LaMnO_3$ $T_x(H_x \sim 200$ kOe, T=0k) and in $CuCl_2 \cdot 2H_2O$ $T_x(H_x=5.5$ kOe, T=0 K), Figs.3 and 4a.

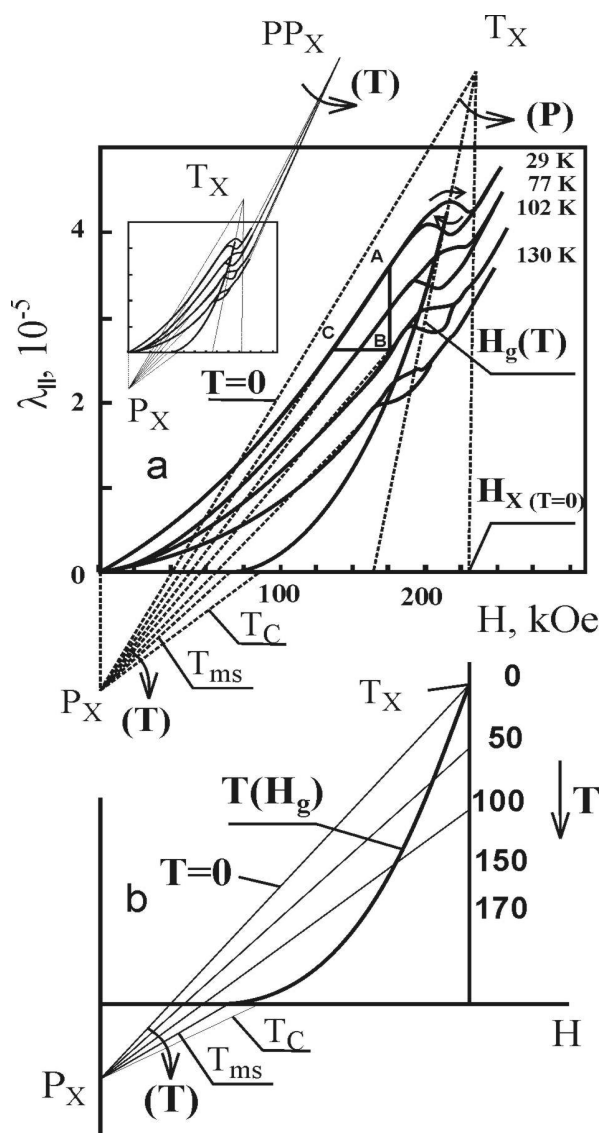

**Figure 3. a) Field dependences of the longitudinal magnetostriction for $LaMnO_3$;**
**b) Field-temperature dependence $T_p(H_g)$ of PT.**

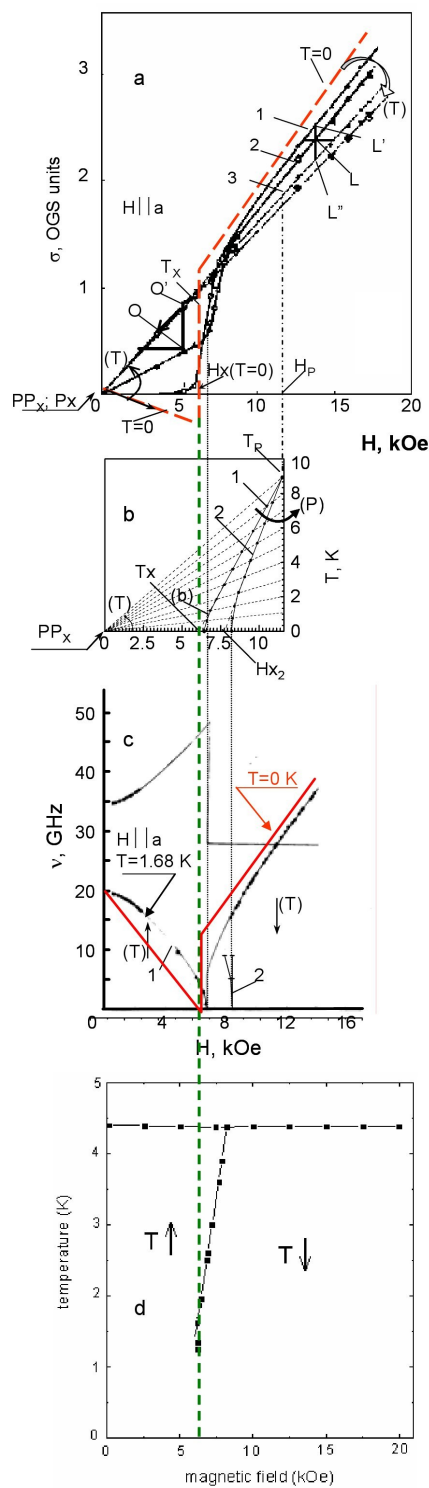

**Figure 4. a) Field dependences of longitudinal magnetostriction for $CuCl_2 \cdot 2H_2O$;**
**b) Field-temperature dependence of changes in the phase transition $T_P(H)$, $T_p$ (H, P) for P = 11,2 kbar in $CuCl_2 \cdot 2H_2O$;**
**c) Dependence of magnetization change in the effect of inverse hysteresis in $CuCl_2 \cdot 2H_2O$;**
**d) Magnetic phase diagram $CuCl_2 \cdot 2H_2O$.**

As there is a problem of obtaining experimental data exactly at 0K, it is very important to establish general signs of the change of properties near this temperature (further we shall call them the secondary signs) that follow the phase transition and are formed under T-P-H influence. The presence of these signs in other systems and samples will define the generalizing role of the mechanisms of elastic deformation stresses in the structural changes and will allow us to single out these regularities in a large number of papers where they are treated by authors as effects, anomalies and peculiarities.

We should add that we pay attention to the relation of elastic magnetic properties and the orientation of the structure with respect to H. Thus, we can find the position of the critical points $P_x$ and $PP_x$ on Figs. 3, 4a. Both magnet-containing and the structural changes connected with the composition and technology of preparation of the samples shift the location of the critical points $P_x$ and $PP_x$ (Fig. 3) and critical lines in the form of temperature-field dependencies of the structural phase transition of type II of $T(H_g)$ that is the boundary of separation of the metallic and semiconducting phases. This is the reason why the location of coincidence of $PP_x$ and $P_x$ (Fig. 4a,c) and the role of sign alteration in the position of the critical line of temperature-field dependence of the structural phase transition $T_p(H)$ show that the critical line separates both the conducting (metallic) phase and the phase where T and H influence the magnetization analogously. The result of thermoelastic compression as a regularity of antiphase thermal influence reflects both the peculiarities of the structure and the anisotropy of elastic deformation stresses. This fact plays the causing role in formation of visible changes of the properties in the form of the jump in the region of the structural phase transition at 0K. We should note that the discussed regularities found in one model structure are repeated in a number of multicomponent systems where the mechanisms of the influence are analogous and follow these conditions of formation of structural phase transitions and the evolution of the properties.

We should say that the consideration of the T and H effects through cooling- heating effect of EAD stresses with the account for the role of the selected critical lines $T_p(H)$ and points $P_x$, $PP_x$ gives a possibility to establish the mechanisms of the change of properties in direct and reverse hysteresis in the course of the analysis of experimental results.

When considering the direct hysteresis (Fig. 3) in magnetostriction dependencies in the area of the phase transition in $LaMnO_3$, we paid attention to the value of thermomagnetic elastic deformation stresses in the influence of thermodynamic parameters T and H [3]. In these papers, the regularity of the influence of P and H through the cooling-heating effects and sign alteration is illustrated as well as the change of priority values of thermal and magnetic elastic deformation stresses in the changes of structural PT, magnetic and resistive properties. We have found the correspondence between the evolution of magnetization in phase states (diagrams) of low-temperature dielectric $CuCl_2 \cdot 2H_2O$ [9].

These results and the position of the selected critical lines and points in the studied properties of $LaMnO_3$ put a question about their role in the changes of hysteresis near the structural PT.

The next example of the change of magnetization in $CuCl_2·2H_2O$ (Fig. 4a,c) allowed us to establish a regularity corresponding to the effect of reverse hysteresis and to define it accounting for the locations of the critical points $PP_x$ and $P_x$ and the changes of temperature – field dependence $T_p(h)$ of the structural PT. The result shows that here the critical line $T_p(H)$ separating the phases is attached to the position of $PP_x$ (Fig. 4b) and the change of the properties (Fig. 4a) at the initial stage of the magnetization with the single-value influence of T and H is the consequence of the cooling effect of the magnetic field and fixes T changes relatively to the critical point $PP_x$ (Fig. 4c).

The further evolution into another phase where the influence of T and H is antiphase fixes the jump of magnetization with the increase of H (Fig. 4a,b) in metallic phase and relatively to $P_x$ (Fig. 4c). Consequently, at the succeeding taking down of magnetic field H and at the rate of the elastic expansion, the heating effect is revealed on the magnetization relatively to $P_x$ with the transition to another phase with respect to $PP_x$. Thus, the properties are under single-valued influence of both T and H. This result is widely spread and follows to the established regularities.

This conclusion is an example of a definition of the well-studied effect of the direct and the reverse hysteresis that underlines the main reason of cooling-heating effect in the change of properties near the structural PT at rigid orientation and anchor to the position of the critical points $PP_x$ and $P_x$. Such treatment is a conclusive proof of the role of mechanisms of elastic deformation stresses in realization of the studied differences and can be an example of the presence of cooling-heating effects and the secondary signs proving the causing role of elastic deformation stresses in the regularities of the direct and reverse hysteresis.

*X-ray experiment results.*

To estimate the elastic parameters and regularities of a single crystal, the influence of the uniform hydrostatic pressure on the parameters of major crystallographic axes was studied. The measurements were done at room temperature in monocrystal $CuCl_2·2H_2O$. The experiment was carried out using X-ray diffractometer at 0-2 kbar. The evaluation of compressibility in Co-radiation yielded the interfacial parameters in a linear form (Fig.5). The linear changes of the parameters and the volume compressibility are obvious proofs of the role of elasticity in structural changes [4].

One of the direct methods of registration of structural changes is X-ray method. The registration of the heat sink process can be made by the change of the lattice parameter due to atomic bonds in the system of single crystal $La_{1-x}Sr_xMnO_3$ with the temperature increase [6]. The evolution of the lattice parameter for all the studied compositions is linear (Fig. 6). The linear regularity of the lattice parameter as a function of temperature with an allowance for error is observed in the whole range of the investigation from 373 K up to 1473 K in single crystal $La_{0.875}Sr_{0.125}MnO_3$ at the rate of thermoelasticity. The same paper presents the results of variation of the ratio La and Sr. Linear regularities stay stable within the whole temperature range. Here the jump-like change of the parameters is presented too, being determined by the presence of SPT in the range of 300-350 K. The transition from rhombohedral symmetry at 88 K to orthorhombic responses at 600 K is fixed. Thus, the cause of the changes of the lattice parameter is the bulk elasticity within the whole temperature region. It is transformed by the heat sink and gives origin to the mechanisms forming the condition of a PT ant the properties of phase states where magnetic properties are started from T~350 K.

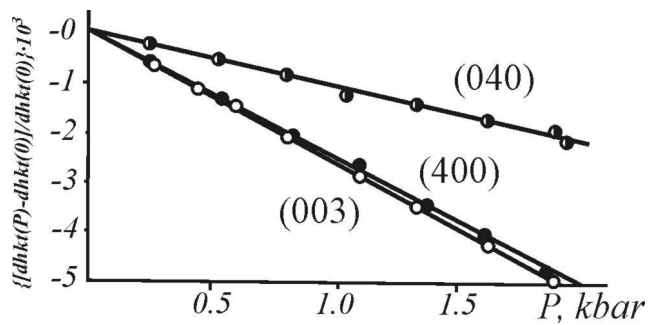

**Figure 5. Dependence of interplanar spacings $d_{400}$, $d_{040}$, $d_{003}$ on uniform compression in $CuCl_2·2H_2O$ [4].**

Analogous X-ray measurements of polycrystal $LaMnO_3$ and $La_{0.7}Sr_{0.3}MnO_3$ have also shown the singled out linear elastic dependencies of the parameters in [7]. The presented linear registration stresses the significant role of DS in the bulk elasticity and the processes of heat sink (cooling). Here we should note that the magnetic

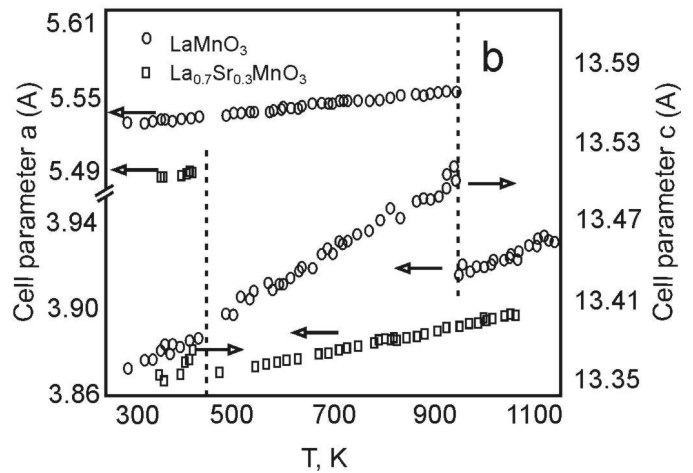

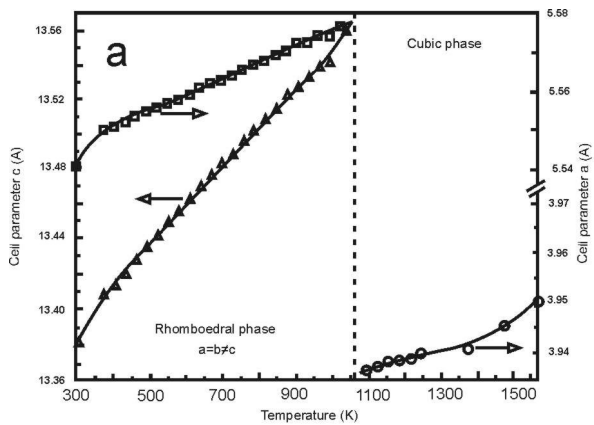

**Figure 6. a) The stability of linear changes in $La_{0.7}Sr_{0.3}MnO_3$ in whole temperature range 350 – 1500 K.**
**b) Monotonous linear increase of crystal lattice parameter of $LaMnO_3$ and $La_{0.7}Sr_{0.3}MnO_3$ polycrystalline samples [7].**

properties are formed below 350 K in these systems (Fig. 6).

*Structure-magnetization correlation in $MnF_2$.*

Let us now to consider the experiments with ultrasound adsorption in the region of PT [8] and to analyze the magnetic properties by using the phase diagram of $MnF_2$ (Fig. 7a). On Fig. 7b, the considered $T_p(H)$ dependence is a field-temperature curve with the critical points $PP_x$ and the temperature parameter evolution corresponding to the regularity of the first-order phase transition for $T_{cm}(T_x=0$ K, $H_x$~90 kOe) and of the phase transition changes with the plus sign. The anomalies of the ultrasound absorption observed near this dependence of PT can be related to the critical region separating phase states.

It has been detected that for H=93 kOe directed along the axis of the symmetry, the magnetostriction has demonstrated a jump of the crystal size [8], a change of the length in high H that is also typical of the first-order structural phase transition. These studies help in the estimation of the magnetoelastic constants for $MnF_2$: $A_0=8.6 \cdot 10^{-13} cm^2/erg$, $A_1=2.1 \cdot 10^{-13} cm^2/erg$, $A_2=1.07 \cdot 10^{-13} cm^2/erg$.

Investigations of the region where the structural phase transition is formed impose demands upon experimental methods. Broad temperature range is implied starting from 4.2 K and lower. For $MnF_2$, the magnetic fields are of $10^5$-$10^6$ Oe with high homogeneity of the fields, high accuracy of the orientation of the external field with respect to crystallographic axes as well as the detailed study of the field-temperature dependence of the resonance absorption in the vicinity of PT.

The explicit jumps of the magnetization properties (Fig. 7c) let us to assume that in $MnF_2$, the first-order structural phase transition restricts the phase states, changes the volume and the symmetry of each phase. This results in differences in the magnetization as well as in changes of the direction of magnetic non-compensation along crystallographic axes (different specific magnetizations) of neighbor phases relatively to changes of the crystal symmetry.

The magnetization jumps are natural factors of the formation of the structural phase transition.

The considered investigations make us to conclude that the phenomenon of jump-like magnetization change is a typical sign of physical processes of T, H effect under which the PT is formed through the mechanisms of elasticity, magnetic elasticity and important elastic and

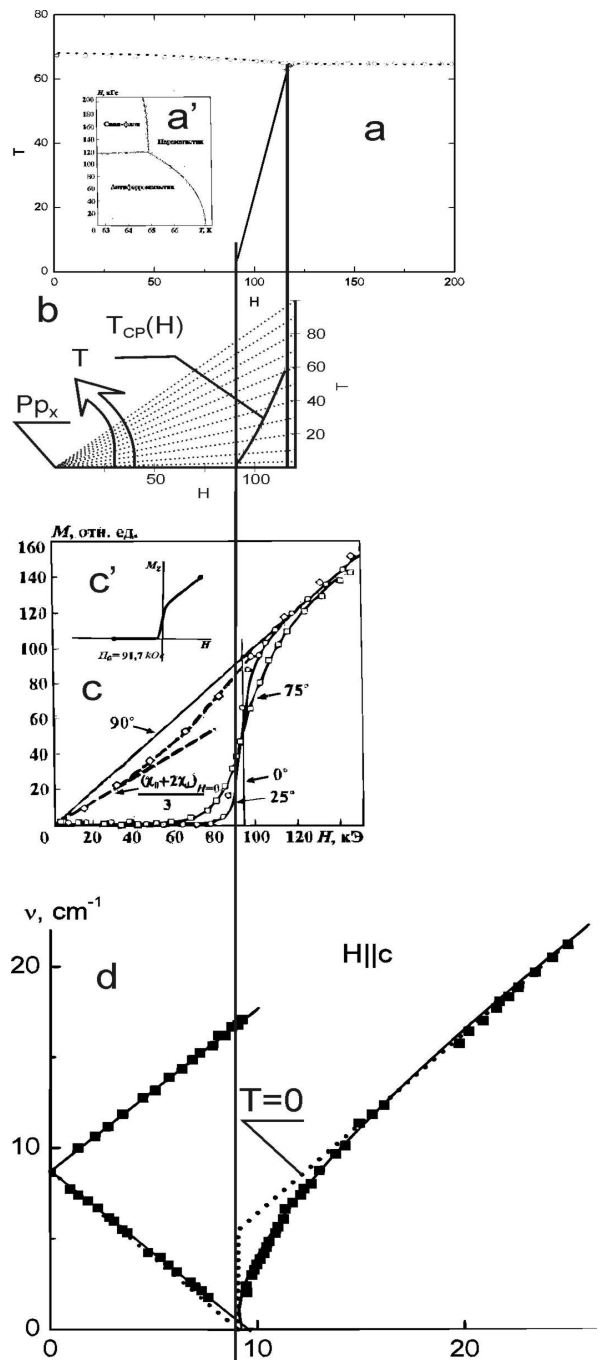

**Figure 7. a, a') Magnetic phase diagram $MnF_2$ in the field parallel to (001) axis.**

**b) Temperature and field dependence of a structural phase transition of the first order with separated critical point $PP_x$ and $T_{pp}$ ($T_x=0$, $H_x\sim93$ kOe).**

**c, c') The jump of magnetization of $MnF_2$ at varied magnetic field orientations with respect to (001) axis at T=4.2 K.**

**d) Frequency and field dependence of resonance absorption of $MnF_2$ in the field parallel to (001) axis at T=4.2 K.**

magnetoelastic anisotropy. The jump-like magnetization change is a sudden change of the properties in a narrow range of fields H~670 Oe. On the background of repeated jumps, there is a region of angles of the field deviation with respect to crystal symmetry, to within ψ~30′. This is shown by magnetization properties on the oscillogramm of Fig. 7b.

The studied phase states differ by the degree of magnetization, first of all. The experimentally observed changes are characterized by different intensity of magnetic non-compensation and magnetization jump in $MnF_2$ for $H_c$=92 kOe makes 100 units of CG-SM. The inhomogeneity of the phases in the course of PT realization is revealed by the critical range of fields and angles. While considering and drawing analogies with formation of the first-order structural phase transition in $CuCl_2·2H_2O$ and $MnF_2$, let us now to analyze the field-frequency resonance absorption branch [8]. For $MnF_2$, the field-frequency dependence was realized in the temperature range of 4.2 K with high-accurate magnetic field orientation along the tetragonal crystal axis for frequencies of 3.5-16.5 $cm^{-1}$. For two single crystals of highly accurate orientation, the resonance absorption decreases the intensity of the second absorption line to the level of resolution near the PT. At low angles of deviation, the absorption was restored. For $MnF_2$, the fields of Pt region correspond to H~0.1+0.005T. The investigated samples were in the form of plates, L=3.5 mm and h=0.07-1.5 mm [8].

Form the above results, it can be concluded, that the regularity of the formation of the jump of the properties under the first-order structural phase transition is common of magnetodielectrics having different composition and compressibility factors. The results obtained for $MnF_2$ are observed for another systems of different compositions still their properties are varied in a similar way [8]. The magnetic non-compensation is maximal along the principal crystallographic axis c, which is relatively small and for the transition to be realized at the point of 0 k, high magnetic field H=90 kOe is necessary (see the dotted line on Fig. 7d).

The crystal and magnetic structure of $MnF_2$ as well as the elastic deformation of the crystal lattice in high magnetic fields are related to the magnetostriction through the mechanisms of magnetoelastic influences under magnetization of the crystal and this is the causal basis for the changes of the properties, which are generalized by regularity of the bulk elasticity effect.

Another confirmation of the results on determination of the region of the structural phase transition in $MnF_2$ is work [8] where the magnetic susceptibility was studied using the samples of different shape. The specific run of the magnetization curves in the region of the jump of the properties for H~92 kOe and T=4.2 K is noted in the initial section. In low fields, the magnetization is low. Then, there follows a jump and a considerable change of the properties (Fig. 7c). Here we remark the analogies of the results. A correspondence to the magnetization jump in $CuCl_2·2H_2O$ is present in lower fields and another temperatures (see Fig. 4).

It follows that as dT/dH, $T_p(H)$ regularity is of positive sign and in the both cases we have the first-order phase transition. Thus, we get $T_p(T_x=0, H_x=90$ kOe) for $MnF_2$ and $T_k(T=0, H_x=5.5$ kOe) for $CuCl_2·2H_2O$. It is also seen that for those single crystals the compared coefficients of compressibility differ by an order of magnitude, but there are regularities of H change in higher fields and PT dynamics corresponds to the "cooling-heating" effect that is an increase of H resulting in T increase at $T_p(H)$, and on the contrary, the reduction of H decreases T. Note that single crystal orientation and the sample shape are in direct correlation with the physical process of the structure change at the expense of elasticity anisotropy restricting the region of PT formation to 0.4kOe for a cylinder and 0.9 kOe for a disc.

Form the analysis of the field-temperature dependences for $MnF_2$ (Fig. 7c) and approximation thereof to 0 k we can distinguish the frequency – independent region which has not been determined in the low-temperature range for 4.2 K. With the highly accurate orientation of the magnetic field, the same was observed and characterized by the position of the structural phase transition in $CuCl_2 \cdot 2H_2O$ single crystal (Fig. 4). Thus, it can be concluded that from the results obtained for alloys based on 3d and 4d tetragonal structure elements, the role of mechanisms of elastically deforming stresses and the anisotropy of elasticity is the determining one with respect to the physical properties of the volume change. The result confirming the analogies with the regular change of the properties on Fig. 7a,b,c,d for $MnF_2$ and Fig. 4a,b,c for $CuCl_2 \cdot 2H_2O$ is the identity of the field-temperature dependences for the first-order PT, $T_p$ with the distinguished critical point $PP_x$ and of the field – frequency dependence with the change at 0 K shown by the dotted line as well as with the region that fixes the positions of PT and phase states.

While separating the changes of the magnetic properties under magneto-elastic effect in the superconducting phase state and drawing analogies with the alternating mechanism of hydrostatic pressure effect on the structure inhomogeneity and PT character, it is possible to relate the regions of the superconducting phase state and magnetism with the processes of the changes of the volume and the symmetry.

**4. Conclusion.**

The analysis of the results of the changes of the field dependencies is concentrated on magneto-elastic, magnetic, electronic, structural interactions connected with the structure rearrangement at the expense of the stresses created by the magnetic field through the mechanisms of magnetic elasticity, which affect both the structural changes and the processes of formation of the region of the first- and second-order structural PT that are recorded by the resonance methods in high magnetic fields and under the hydrostatic pressure. Showing the experimental results for real physical processes, we try to attract the attention of researchers stating that theoretical models are needed taking into account the energies of elastically deforming stresses and applied to a practical case. The analysis and the methods used to ground the analogies in the study of magneto-dielectrics and superconductors enable us to generalize and to connect the regularities of magnetostriction and structural changes occurring during the formation of the first-order structural phase transition. As a consequence, the changes of magnetization and magnetostriction are in accordance with the general regularities of magneto-elasticity that is a part of the total bulk elasticity.

**References**

[1] P.L. Kapitsa 1988 *Strong magnetic fields* ( Moscow: Sci. Works. Science)

[2] A.M. Kadomtseva, Yu.V. Popov, G.P. Vorob'ev, K.I. Kamilov, V.Yu. Ivanov, A.A. Muhin and A.M. Balbashov 2000 *Phys. Tverd. Tela.* **42** 1077

[3] P.I. Polyakov, S.S. Kucherenko 2002 *JMMM* **248** 396
P.I. Polyakov, S.S. Kucherenko 2004 *JMMM* **278** 138

[4] A.A. Galkin, V.G. Baryahtar, S.V. Ivanova, P.I. Polyakov 1979 *Fiz. Tverd. Tela.* **5** 1515
A.A. Galkin, V.G. Baryahtar, S.V. Ivanova, P.I. Polyakov 1979 *Fiz. Tverd. Tela.* **9** 2580

[5] T. Timusk and D. B. Tanner 1989 *Physical Properties of High Temperature Superconductors /* Edited by D. M. Ginsberg ( Singapore: World Scientific)

[6] R.F.C. Marques, M. Jafelicci Ir., C.O. Paiva-Sautor, L.C. Varanda, R.H.M. Godoi 2001 *JMMM* **226** 812

[7] V.A. Turchenko, P.I. Polyakov 2007 *ICFM-2007*, Ukraine, Partenit, October 3-8, 188
V.A. Turchenko, P.I. Polyakov 2007, *FTT-2007* Minsk, October 26-29 184

[8] V.V. Eremenko, V.A. Sirenko 2004 *Magnetic and magnetoelastic properties of antiferromagnets and superconductors* ( Kiev: Naukova Dumka)

[9] P.I. Polyakov, T.A. Ryumshyna 2009 *Magnetism and Laws of Bulk Elasticity* (Kerala, India: TransWorld Research Network)